\documentclass[11pt, journal, draftcls, onecolumn]{IEEEtran}
\usepackage{amsmath}
\usepackage{nicefrac}
\usepackage{amsfonts}
\usepackage{multirow}
\usepackage{multicol}
\usepackage{amssymb}
\usepackage[normalem]{ulem}
\usepackage{color}
\usepackage{graphicx}
\usepackage[capitalize,nameinlink]{cleveref}
\usepackage[margin=0.5in]{geometry}
\usepackage[numbers,square, sort&compress]{natbib}
\newtheorem{lemma}{\textit{Lemma}}
\newtheorem{cor}{\textit{Corollary}}
\newtheorem{theorem}{\textit{Theorem}}
\newtheorem{proposition}{\textit{Proposition}}
\newenvironment{remark}[1][Remark]{\begin{trivlist}
\item[\hskip \labelsep {\bfseries #1}]}{\end{trivlist}}

\begin{document}


\title {Are Guessing, Source Coding, and  Tasks Partitioning Birds of a Feather?}

\author{\IEEEauthorblockN{M. Ashok Kumar\IEEEauthorrefmark{1}, Albert Sunny\IEEEauthorrefmark{1}, Ashish Thakre\IEEEauthorrefmark{2}, Ashisha Kumar\IEEEauthorrefmark{2}, G. Dinesh Manohar\IEEEauthorrefmark{3}}
		
\thanks{\IEEEauthorrefmark{1}Indian Institute of Technology Palakkad, Kerala, India
Email: \{ashokm, albert\}@iitpkd.ac.in}
\thanks{\IEEEauthorrefmark{2}Indian Institute of Technology Indore, Madhya Pradesh, India
	Email: ashishblthakre10@gmail.com, akumar@iiti.ac.in}
	
\thanks{\IEEEauthorrefmark{3}Technical University of Munich, Bavaria, Germany
Email: gdineshnathan@yahoo.com}
}
\maketitle

\begin{abstract}
This paper establishes a close relationship among the four information theoretic problems, namely Campbell’s source coding, Ar{\i}kan’s guessing, Huleihel et al. 's memoryless  guessing and  Bunte  and  Lapidoth’s tasks partitioning problems. We first show that the aforementioned problems are mathematically related via a general moment minimization problem whose optimum solution is given in terms of R\'enyi Entropy. We then propose a general framework for the mismatched version of these problems and establish all the asymptotic results using this framework. Further, we study an ordered tasks partitioning problem that
turns out to be a generalization of Ar{\i}kan's guessing problem. Finally, with the help of this general framework, we establish an equivalence among all these problems, in the sense that, knowing an asymptotically optimal solution in one problem helps us find the same in all other problems.
\end{abstract}

\section{Introduction}
\label{sec:introduction}
The concept of entropy is very central to information theory. For example, number of bits required on average (per letter) to compress a source with (finite) alphabet set $\mathcal{X}$ and probability distribution $P$ is the \emph{Shannon entropy} $H(P)$. If the compressor does not know the true distribution $P$, but assumes a distribution $Q$ (mismatch), then the number of bits required for compression is $H(P) + I(P,Q)$, where $I(P,Q)$ is the \emph{entropy of $P$ relative to $Q$} (or the Kullback-Leibler divergence). In his seminal paper, Shannon \cite{Shannon} argued that $H(P)$ can also be regarded as a measure of uncertainty. Subsequently, R\'enyi \cite{Renyi} introduced an alternate measure of uncertainty, now known as {\em R\'enyi entropy} of order $\alpha$, as 
\begin{align}
\label{eq:renyi_entropy}
H_{\alpha}(P):=\frac{1}{1-\alpha}\log \sum \nolimits_{x \in \mathcal{X}} P(x)^{\alpha},
\end{align}
where $\alpha >0$ and $\alpha \neq 1$. R\'enyi entropy can also be regarded as a generalization of the Shannon entropy as $\lim_{\alpha\to1}H_{\alpha}(P) = H(P)$. Refer Aczel and Daroczy \cite{Aczel} and the references therein for an extensive study of various measures of uncertainty and their characterizations.

In 1965, Campbell \cite{Campbell} gave an operational meaning to R\'enyi entropy. He showed that if one minimizes the cumulants of code lengths, then the optimal cumulant is R\'enyi entropy $H_{\alpha}(P)$. He also showed that the optimal cumulant can be achieved by encoding sufficiently long sequences of symbols. Sundaresan \cite[Th. 8]{Sundaresan} (c.f. Blumer and McElice \cite{BlumerMcElice}) showed that in the mismatched case, the optimal cumulant is $H_{\alpha}(P) + I_{\alpha}(P,Q)$, where 
\begin{align}
I_\alpha (P,Q) := 
\frac{\alpha}{1-\alpha}{\rm log}\Big ( \sum\nolimits_ {x \in \mathcal{X}} P(x)
\Big [ & \sum\nolimits_ {x' \in \mathcal{X}}
\Big ( \frac{Q(x')}{Q(x)}\Big )^\alpha \Big ]^\frac{1- \alpha}{\alpha}\Big ) - H_\alpha(P), \label{eq:sundaresan_divergence}
\end{align}

\noindent
is {\em $\alpha$-entropy of $P$ relative to $Q$} or {\em Sundaresan's divergence} \cite{Sundaresan02ISIT}. $I_{\alpha}(P,Q)\ge 0$ with equality if and only if $P=Q$. Hence $I_{\alpha}(P,Q)$ can be interpreted as the penalty for not knowing the true distribution. $I_{\alpha}$-divergence can also be regarded as a generalization of the Kullback-Leibler divergence as $\lim_{\alpha\to 1}I_{\alpha}(P,Q) = I(P,Q)$. Refer \cite{Sundaresan, KumarS15J1} for detailed discussions on the properties of $I_\alpha$. Lutwak et al. also independently identified $I_{\alpha}$ in the context of maximum R\'enyi entropy and called it {\em $\alpha$-R\'enyi relative entropy} \cite[Eq.~(4)]{200501TIT_LutYanZha}. $I_{\alpha}$, for $\alpha >1$, also arises in robust inference problems (see \cite{KumarS15J2} and the references therein).

In \cite{Massey}, Massey studied a guessing problem where one is interested in the expected number of guesses required to guess a random variable $X$ that assumes values from an infinite set, and found a lower bound in terms of Shannon entropy. Ar{\i}kan \cite{Arikan} studied a guessing problem on finite alphabet sets and showed that R\'enyi entropy arises as the optimal solution when minimizing moments of number of guesses. Subsequently, Sundaresan \cite{Sundaresan} showed that the penalty in guessing according to a distribution $Q$ when the true distribution is $P$, is given by $I_\alpha(P,Q)$. Bunte and Lapidoth \cite{Bunte} studied a problem of partitioning  tasks and showed that R\'enyi entropy and Sundaresan's divergence play a similar role in the optimal number of tasks performed. Quite recently, Huleihel et al. \cite{Huleihel} studied the memoryless guessing problem, a variant of Ar{\i}kan's guessing problem, with i.i.d. (independent and identically distributed) guesses and showed that the minimum attainable factorial moments of number of guesses is the R\'enyi entropy.

\subsection{Our Contribution}
On studying the aforementioned four problems namely, Campbell’s source coding, Ar{\i}kan’s guessing, Huleihel et al.’s memoryless guessing, and Bunte and Lapidoth’s tasks partitioning, we observe a close relationship among them. In all these problems, the objective is to minimize moments or factorial moments of random variables, and R\'enyi entropy and Sundaresan's divergence arise in optimal solutions. This motivated us to seek out mathematical and operational commonalities that exist in these problems. 
The main contributions of this paper are as follows.

\begin{itemize}
    \item A general framework for the problems on source coding, guessing and tasks partitioning.
    
    
    \item A unified approach to derive bounds for the mismatched version of these problems.
    
    \item A generalized tasks partitioning problem.
    
    \item Establishing operational commonality among the problems.
\end{itemize}

The remainder of the paper is organized as follows. In \cref{sec:unified_approach}, we present our unified framework, and find conditions under which lower and upper bounds are attained. In \cref{sec:problem_statements}, we present four well-known information-theoretic problems, namely,  Campbell’s source coding, Ar{\i}kan’s guessing, Huleihel et al.'s memoryless guessing, and Bunte and Lapidoth’s tasks partitioning, and re-establish and refine major results pertaining to these problems. In \cref{sec:generalized_task}, we present and solve a generalized tasks partitioning problem. In \cref{sec:connection}, we establish a connection among the aforementioned problems. Finally, we summarize and conclude the paper in \cref{sec:summary}. 

\section{A General Minimization Problem} \label{sec:unified_approach}

In this section, we present a general minimization problem whose optimum solution evaluates to R\'enyi entropy. We will later show that all problems stated in \cref{sec:problem_statements} are particular instances of this general problem.

\begin{proposition} \label{prop:lower_bound_unified}
Let $\psi: \mathcal{X} \to (0,\infty)$ be such that $\sum_{x \in \mathcal{X}} \psi(x)^{-1} \le k$ for some $k>0$. Then, for $\rho\in (-1,0)\cup (0,\infty)$,  
\begin{equation}
\label{eq:lower_bound_on_cumulant}
\frac{1}{\rho} \log \mathbb{E}_{P}[\psi(X)^{\rho}] \geq  H_{\alpha}(P) - \log k, 
\end{equation}
where $\mathbb{E}_P[\cdot]$ denotes the expectation with respect to probability distribution $P$ on $\mathcal{X}$, $H_{\alpha}(P)$ is the R\'enyi entropy of order $\alpha$, and $\alpha := \alpha(\rho) = 1/(1+\rho)$. The lower bound is achieved if and only if
\begin{equation}
\label{eqn:lower_bound_attain}
    \psi(x)^{-1} =  {k \cdot P(x)^\alpha}/{Z_{P,\alpha}} \quad \text{for } x \in \mathcal{X},
\end{equation}
 where $Z_{P,\alpha} := \sum_{x\in\mathcal{X}}P(x)^\alpha$. 
\end{proposition}

\begin{IEEEproof}
    Observe that
    \begin{eqnarray*}
    \textrm{sgn}(\rho) \sum \nolimits_{x \in \mathcal{X}}  P(x) \psi(x)^{\rho} & = & \textrm{sgn}(\rho) \sum \nolimits_{x \in \mathcal{X}}  P(x)^{\alpha} \left(\frac{\psi(x)^{-1}}{P(x)^{\alpha}}\right)^{-\rho}\\
    & \stackrel{(a)}{\ge} & \textrm{sgn}(\rho) \Big(\sum_x  P(x)^{\alpha}\Big)\cdot \left(\frac{\sum_x \psi(x)^{-1}}{\sum_x P(x)^{\alpha}}\right)^{-\rho}\\
    & = & \textrm{sgn}(\rho) \Big(\sum_x  P(x)^{\alpha}\Big)^{1+\rho}\cdot \Big(\sum_x \psi(x)^{-1}\Big)^{-\rho}\\
    & \stackrel{(b)}{\ge} & \textrm{sgn}(\rho) \Big(\sum_x  P(x)^{\alpha}\Big)^{1+\rho}\cdot k^{-\rho},
    \end{eqnarray*}
where $(a)$ is due to the generalised log-sum inequality \cite[Eq.~(4.1)]{2004xxFTCIT_Csi_Shi} applied to the function $f(x) = \textrm{sgn} (\rho) \cdot x^{-\rho}$; and $(b)$ follows from the hypothesis 
$\sum_{x} \psi(x)^{-1}$ $ \le k$. By taking $\log$ and then dividing by $\rho$, we get \eqref{eq:lower_bound_on_cumulant}. Equality holds in $(a)$ if and only if $\psi(x)^{-1} = \nu P(x)^{\alpha}$ for some constant $\nu$ and in (b) if and only if $\sum_{x} \psi(x)^{-1} = k$. These conditions are met only when $\psi(x)$ satisfies \eqref{eqn:lower_bound_attain}.
\end{IEEEproof}

The left-side of \eqref{eq:lower_bound_on_cumulant} is called  {\em normalised cumulant of $\psi(X)$ of order $\rho$}. The measure $P^{(\alpha)}(x) := {P(x)^\alpha}/{Z_{P,\alpha}}$ in \eqref{eqn:lower_bound_attain} that attains the lower bound is called {\em $\alpha$-scaled measure} or {\em escort measure} of $P$. This measure also arises in robust inference \cite[Eq.~(7)]{KumarS15J2} and statistical physics \cite{TsallisMP98}.
The above proposition can also be proved using a  variational formula as follows. By a version of Donsker-Varadhan variational formula \cite[Propostion~4.5.1]{1997Paul}, for any real-valued $f$ on $\mathcal{X}$, we have
\begin{equation}
\label{eqn:variational_formula}
    \log \mathbb{E}_P[2^{f(X)}] = \max_Q \{\mathbb{E}_Q[f(X)] - D(Q\|P)\},
\end{equation}
where the $\max$ is over all probability distributions $Q$ on $\mathcal{X}$. Taking $\rho >0$ and $f(x) = \rho \log \psi(x)$ in \eqref{eqn:variational_formula} we have
\begin{eqnarray*}
\log \mathbb{E}_P[\psi(X)^\rho] & = & \max_Q \{\rho \mathbb{E}_Q[\log\psi(X)] - D(Q\|P)\}\\
& = & \max_Q \Big\{-\rho \sum_x Q(x)\log\frac{\psi(x)^{-1}Q(x)}{Q(x)} - D(Q\|P)\Big\}\\
& = & \max_Q \Big\{\rho H(Q) -\rho \sum_x Q(x)\log\frac{\psi(x)^{-1}}{Q(x)} - D(Q\|P)\Big\}\\
& \stackrel{(a)}{\ge} & \max_Q \Big\{\rho H(Q) -\rho \log\sum_x\psi(x)^{-1} - D(Q\|P)\Big\}\\
& \stackrel{(b)}{\geq} & \rho \max_Q \Big\{H(Q) - \frac{1}{\rho} D(Q\|P)\Big\}  -\rho \log k,
\end{eqnarray*}
where (a) is by the log-sum inequality \cite[Eq.~(4.1)]{2004xxFTCIT_Csi_Shi} and (b) is by applying the constraint $\sum_{x} \psi(x)^{-1} \leq k$. For $\rho\in (-1,0)$, the inequalities in (a)  and (b) are reversed and the last $\max$ is replaced by $\min$. Hence, (\ref{eq:lower_bound_on_cumulant}) follows as the last $\max$ is equal to $H_\alpha(P)$ by \cite[Th.~1]{Sha2011ISIT}. Equality in $(a)$ and $(b)$ holds if and only if $\psi(x)^{-1} = k\cdot Q(x)$. Also the last max is attained when $Q(x) = {P(x)^\alpha}/{Z_{P,\alpha}} \text{ for } x \in \mathcal{X}$. This completes the proof. \hfill\QED

\noindent
The following is the analogous one for Shannon entropy.


\begin{proposition}
 \label{prop:lower_bound_logfunc}
Let $\psi: \mathcal{X} \to (0,\infty)$ be such that $\sum_{x \in \mathcal{X}} \psi(x)^{-1} \leq k$, then 
\begin{align}
\label{eq:lower_bd_exp_length}
    \mathbb{E}_P\left[\log {\psi(X)} \right] \geq  H(P) - \log k.
\end{align}
The lower bound is achieved if and only if $\psi(x)^{-1} = k\cdot P(x) \,\, \forall x \in \mathcal{X}$. 
\end{proposition}
\begin{IEEEproof}
\begin{eqnarray*}
    \mathbb{E}_P\left[\log {\psi(X)} \right] & = & - \sum_x P(x)\log\frac{P(x)\cdot\psi(x)^{-1}}{P(x)}\\
    & = & H(P) - \sum_x P(x)\log\frac{\psi(x)^{-1}}{P(x)}\\
    & \ge & H(P) - \log\sum_x \psi(x)^{-1}\\
    & \ge & H(P) - \log k,
\end{eqnarray*}
where the pen-ultimate inequality is due to the log-sum inequality. Equality holds in both inequalities if and only if $\psi(x)^{-1} = k\cdot P(x) \,\, \forall x \in \mathcal{X}$.
\end{IEEEproof}

It is interesting to note that $\frac{1}{\rho} \log \mathbb{E}_P[\psi(X)^{\rho}] \to \mathbb{E}_P[\log ({\psi(X)})]$ and $H_{\alpha}(P) \to H(P)$ as $\rho\to 0$ in \eqref{eq:lower_bound_on_cumulant}.
We now extend Propositions \ref{prop:lower_bound_unified} and \ref{prop:lower_bound_logfunc} to sequences of random variables. Let $\mathcal{X}^n$ be the set of all $n$-length sequences of elements of $\mathcal{X}$, and $P_n$ be the $n$-fold product distribution of $P$ on $\mathcal{X}^n$, that is, for $x^n := x_1,\dots, x_n\in \mathcal{X}^n$, $P_n(x^n) = \prod_{i=1}^n P(x_i)$.

\begin{cor} \label{corr:sequence_bound_lower}
Given any $n \geq 1$, if $\psi_n: \mathcal{X}^{n} \to [0,\infty)$ is such that $\sum_{x^n \in \mathcal{X}^n} \psi_n(x^n)^{-1}$  
$\leq k_n$ for some $k_n >0$, then
\begin{enumerate}
    \item[(a)] For $\rho \in (-1,0) \cup (0, \infty)$, 
\[
\liminf_{n \to \infty} \frac{1}{n\rho} \log \mathbb{E}_{P_n}[\psi_n(X^n)^{\rho}] \geq H_{\alpha}(P)  - \limsup_{n \to \infty}\frac{\log k_n}{n}.
\]
	\item[(b)] 
$$ \liminf_{n \to \infty} \frac{1}{n}  \mathbb{E}_{P_n}[\log \psi_n(X^n)] \geq H(P) - \limsup_{n \to \infty}\frac{\log k_n}{n},$$
\end{enumerate}
where $\mathbb{E}_{P_n}[\cdot]$ denotes the expectation with respect to probability distribution $P_n$ on $\mathcal{X}^n$.
\end{cor}

\begin{IEEEproof}
It is easy to see that $H_{\alpha}(P_n) = n H_{\alpha}(P)$ and $H(P_n) = n H(P)$. Applying Propositions~\ref{prop:lower_bound_unified} and \ref{prop:lower_bound_logfunc}, dividing throughout by $n$ and taking $\liminf n \to \infty$, the results follow.
\end{IEEEproof}

\subsection{A General Framework for Mismatched Cases} \label{subssec:mismatch_case}
In this sub-section, we establish a unified approach for cases when there mismatch between assumed and true distributions.

\begin{proposition}
\label{thm:bound2_unified_mismatch}
	Let $\rho > -1$, $\alpha=1/(1+\rho)$, and $Q$ be a probability distribution on $\mathcal{X}$. For $n \geq 1$, let $Q_n$ be the $n$-fold product distribution of $Q$ on $\mathcal{X}^n$. If $\psi_n: \mathcal{X}^n \to (0,\infty)$ is such that
	\begin{equation}
	\label{eqn:bound_on_psi}
	    \psi_n(x^n) \leq \frac{c_n \cdot  Z_{Q_n,\alpha}}{Q_n(x^n)^{\alpha}}
	\end{equation}
	for some $c_n > 0$, then
	\begin{enumerate}
	    \item[(a)] for $\rho \neq 0$,
	    $$\text{sgn}(\rho) \cdot \mathbb{E}_{P_n}[\psi_n(X^n)^\rho] \leq \text{sgn}(\rho) \cdot 2^{n \rho [H_{\alpha}(P) + I_{\alpha}(P,Q) + n^{-1} \log c_n]},$$
	    
	    \item[(b)] for $\rho \neq 0$,
	    $$ \limsup_{n \to \infty} \frac{1}{n \rho} \log\mathbb{E}_{P_n}[\psi_n(X^n)^{\rho}] \leq  H_{\alpha}(P) + I_{\alpha}(P,Q) + \limsup_{n \to \infty} \frac{1}{n} \log c_n ,$$
	    \item[(c)] for $\rho = 0$,  
	    $$\mathbb{E}_{P_n}[ \log \psi_n(X^n)] \leq {n [H(P) + I(P,Q) + n^{-1} \log c_n]},$$
	    \item[(d)] for $\rho = 0$,
	    $$ \limsup_{n \to \infty} \frac{1}{n} \mathbb{E}_{P_n}[\log \psi_n(X^n)] \leq  H(P) + I(P,Q) + \limsup_{n \to \infty} \frac{1}{n} \log c_n.$$
	\end{enumerate}
\end{proposition}
\begin{IEEEproof}
\textit{Part (a):}  From \eqref{eqn:bound_on_psi}, we have
\begin{align}
\text{sgn}(\rho) \cdot \mathbb{E}_{P_n}[\psi_n(X^n)^{\rho}] &= \text{sgn}(\rho) \cdot \sum \nolimits_{x^n \in \mathcal{X}^n} P_n(x^n) \psi_n(x^n)^{\rho} \nonumber \\
& \leq \text{sgn}(\rho) \cdot c_n^{\rho} Z_{Q_n,\alpha}^{\rho} \sum \nolimits_{x^n \in \mathcal{X}^n} P_n(x^n) Q_n(x^n)^{-\alpha\rho} \nonumber \\
& = \text{sgn}(\rho) \cdot  2^{\rho[H_\alpha(P_n) + I_{\alpha}(P_n,Q_n) + \log c_n]} \nonumber \\
& = \text{sgn}(\rho) \cdot  2^{n \rho[H_\alpha(P) + I_{\alpha}(P,Q) + n^{-1} \log c_n]}, \label{eq:bound2_eq1}
\end{align}
where the penultimate equality holds from the definition of $I_{\alpha}$, and the last one holds because $H_{\alpha}(P_n) = n H_{\alpha}(P)$, $I_{\alpha}(P_n, Q_n) = n I_{\alpha}(P,Q)$.
\medskip

\noindent
\textit{Part (b):} Taking $\log$, dividing throughout by $n \rho $, and then applying $\limsup$ successively on both sides of \eqref{eq:bound2_eq1}, the result follows.
\medskip

\noindent
\textit{Part (c):} When $\rho = 0$, we have $\alpha = 1$ and  \eqref{eqn:bound_on_psi} becomes $\psi_n(x^n) \leq \frac{c_n}{Q_n(x^n)}$. Hence 
\begin{align}
 \mathbb{E}_{P_n}[\log \psi_n(X^n)] &=  \sum \nolimits_{x^n \in \mathcal{X}^n} P_n(x^n) \log \psi_n(x^n) \nonumber \\
& \leq \log c_n +   \sum \nolimits_{x^n \in \mathcal{X}^n} P_n(x^n) \log (1/Q_n(x^n)) \nonumber \\
& = \log c_n + H(P_n) + I(P_n, Q_n) \nonumber \\
& = n(H(P) + I(P,Q) + n^{-1} \log c_n), \label{eq:bound2_eq2}
\end{align}
where the last equality holds because $H(P_n) = n H(P)$, and $I(P_n, Q_n) = n I(P,Q)$.

\noindent
\textit{Part (d):} Dividing \eqref{eq:bound2_eq2} throughout by $n$, and taking limsup on both sides, the result follows.
\end{IEEEproof}

\begin{proposition}
\label{thm:bound3_unified_mismatch}
	Let $\rho > -1$, $\alpha=1/(1+\rho)$, and $Q$ be a probability distribution on $\mathcal{X}$.  For $n \geq 1$, let $Q_n$ be the $n$-fold product distribution of $Q$ on $\mathcal{X}^n$. Suppose $\psi_n: \mathcal{X}^n \to (0,\infty) $ is such that
	\[
	\psi_n(x^n)\geq \frac{a_n ~Z_{Q_n,\alpha}}{Q_n(x^n)^{\alpha}}
	\]
	for some $a_n > 0$, then 
	\begin{enumerate}
	    \item[(a)] For $\rho \neq 0$,
	    \[
	    \text{sgn}(\rho) \cdot \mathbb{E}_{P_n}[\psi_n(X^n)^{\rho}] \geq \text{sgn}(\rho) \cdot 2^{n \rho(H_{\alpha}(P) + I_{\alpha}(P,Q) + n^{-1} \log a_n)},
	    \]
	    \item[(b)] For $\rho \neq 0$,
	    \[
	    \liminf_{n \to \infty} \frac{1}{n \rho} \log\mathbb{E}_{P_n}[\psi_n(X^n)^{\rho}] \geq  H_{\alpha}(P) + I_{\alpha}(P,Q) + \liminf_{n \to \infty} \frac{1}{n} \log a_n,
	    \]
	    \item[(c)] For $\rho = 0$,
	    \[
	    \mathbb{E}_{P_n}[\log\psi_n(X^n)]\geq {n (H(P) + I(P,Q) + n^{-1} \log a_n)},
	    \]
	    \item[(d)] For $\rho = 0$,
	    \[
	    \liminf_{n \to \infty} \frac{1}{n} \mathbb{E}_{P_n}[ \log \psi_n(X^n)]\geq H(P) + I(P,Q) + \liminf_{n \to \infty} \frac{1}{n} \log a_n.
	    \]
	\end{enumerate}
\end{proposition}
\begin{IEEEproof}
Similar to proof of \cref{thm:bound2_unified_mismatch}.
\end{IEEEproof}

\section{Problem Statements and Known Results} \label{sec:problem_statements}
In this section we discuss Campbell's source coding problem, Ar{\i}kan's guessing problem, Huleihel et al.'s memoryless guessing problem, and Bunte and Lapidoth's tasks partitioning problem. Using the general framework presented in the previous section, we re-establish known results, and present a few new results relating to these problems.

\subsection{Source Coding Problem}

Let $X$ be a random variable that assumes values from a finite alphabet set $\mathcal{X} = \{a_1,\ldots,a_m\}$ according to a probability distribution $P$. The tuple $(\mathcal{X},P)$ is usually referred to as a \emph{source}. A {\em binary code} $C$ is a mapping from $\mathcal{X}$ to the set of finite length binary strings.  Let $L(C(X))$ be the length of code $C(X)$. The objective is to find a \emph{uniquely decodable} code that minimizes the expected code-length, that is, 

$$\textrm{Minimize}\quad \mathbb{E}_{P}[L(C(X))]$$ 
over all uniquely decodable codes $C$. Kraft and McMillan independently proved the following relation between uniquely decodable codes and their code-lengths.

\medskip
\noindent
\textbf{Kraft-McMilan Theorem \cite{Cover}}: \textit{If $C$ a uniquely decodable code, then}
\begin{equation}
    \label{eq:KraftMcmilan}
	\sum \nolimits_{x \in \mathcal{X}} 2^{-L(C(x))}\leq 1.
\end{equation}
\textit{Conversely, given a length sequence that satisfies the above inequality, there exists a uniquely decodable code $C$ with the given length sequence.}

Thus, one can confine the search space for $C$ to codes satisfying the Kraft-McMillan inequality \eqref{eq:KraftMcmilan}.
\medskip

\noindent
\textbf{Theorem 5.3.1 of \cite{Cover}}: \textit{
If $C$ is a uniquely decodable code, then $\mathbb{E}_{P}[L(C(X))] \ge H(P)$.}
\begin{IEEEproof}
Choose $\psi(x) = 2^{L(C(x))}$, where $L(C(x))$ is the length of code $C(x)$ assigned to alphabet $x$. Since $C$ is uniquely decodable, from (\ref{eq:KraftMcmilan}), we have $\sum_{x \in \mathcal{X}} \psi(x)^{-1} \leq 1$. Now, an application of \cref{prop:lower_bound_logfunc} with $k=1$ yields the desired result.
\end{IEEEproof}

\begin{theorem}
Let $X^n:= X_1,\ldots,X_n$ be an i.i.d. sequence from $\mathcal{X}^n$ following the product distribution $P_n(X^n) = \prod_{i=1}^n P(X_i)$. Let $Q_n(X^n) = \prod_{i=1}^n Q(X_i)$, where $Q$ is another probability distribution. Let $C_n$ be a code such that $L(C_n(X^n)) = \lceil -\log Q_n(X^n) \rceil$. Then, $C_n$ satisfies Kraft-McMillan inequality and
$$\lim_{n \to \infty}\frac{\mathbb{E}_{P_n}[L(C_n(X^n))]}{n} = H(P) + I(P,Q).$$
\end{theorem}
\begin{IEEEproof}
Choose $\psi_n(x^n) = 2^{L(C_n(x^n))}$, where $L(C_n(x^n)$ is the length of code $C_n(x^n)$ assigned to sequence $x^n$. Then, we have
\begin{align*}
\psi_n(x^n) = 2^{L(C_n(x^n))} = 2^{\lceil - \log Q_n(x^n) \rceil}  \leq 2 \cdot 2^{- \log Q_n(x^n) } = 2/Q_n(x^n).   
\end{align*}
\noindent
An application of \cref{thm:bound2_unified_mismatch} with $c_n = 2$ yields $
\limsup_{n \to \infty}\mathbb{E}_{P_n}[L(C_n(X^n))]/n  \leq  H(P) + I(P,Q)
$. Further, we also have
\begin{align*}
\psi_n(x^n) = 2^{L(C_n(x^n))} = 2^{\lceil - \log Q_n(x^n) \rceil}  \geq 2^{- \log Q_n(x^n)} = 1/Q_n(x^n).
\end{align*}
An application of \cref{thm:bound3_unified_mismatch} with $a_n = 1$ gives $
\liminf_{n \to \infty}\mathbb{E}_{P_n}[L(C_n(X^n))]/n  \geq  H(P) + I(P,Q)$.
\end{IEEEproof}

\subsection{Campbell Coding Problem} \label{sec:campbell_coding}

In Campbell's coding problem, the setup is identical to Shannon's source coding problem. However, instead of minimizing the expected code-length, we are interested in minimizing the normalized cumulant of code lengths, that is,
\begin{align*}
\text{Minimize} \quad\frac{1}{\rho}\log\mathbb{E}_{P}[2^{\rho L(C(X))}],
\end{align*}
over all uniquely decodable codes $C$, and $\rho >0$. A similar problem was considered by Humblet in \cite{Humblet} for minimizing buffer overflow  probability. A lower bound for the normalized cumulants in terms of R\'enyi entropy was provided by Campbell \cite{Campbell}.

\medskip
\noindent
\textbf{Lemma 1 of \cite{Campbell}}:
\textit{Let $C$ be a uniquely decodable code. Then,}
	\begin{align}
	\label{eq:Campbell_lower_bound}
	\frac{1}{\rho}\log\mathbb{E}_{P_n}[2^{\rho L(C(X))}]\geq H_{\alpha}(P),
	\end{align}
	\textit{where $\alpha = {1}/{(1+\rho)}.$}
\begin{IEEEproof}
Apply \cref{prop:lower_bound_unified} with $\psi(x) = 2^{L(C(x))}$ and $k=1$.
\end{IEEEproof}

\noindent
Notice that, if we ignore the integer constraint of the length function, then
\begin{align}
	\label{eq:generalized_code_lengths}
	L(C(x))= \log  \frac{Z_{P,\alpha}}{P(x)^\alpha},
	\end{align}
	with $Z_{P, \alpha}$ as in \cref{prop:lower_bound_unified}, satisfies \eqref{eq:KraftMcmilan} and achieves the lower bound in \eqref{eq:Campbell_lower_bound}. Campbell also showed that the lower bound in \eqref{eq:Campbell_lower_bound} can be achieved by encoding long sequences of symbols with code-lengths close to \eqref{eq:generalized_code_lengths}.

\medskip
\noindent
\textbf{Theorem 1 of \cite{Campbell}}:
\textit{If $C_n$ is a uniquely decodable code such that}
\begin{align} \label{eq:optimal_code_length_campbell}
L(C_n(x^n))= \Big\lceil\log  \frac{Z_{P_n,\alpha}}{P_n(x^n)^\alpha}\Big\rceil,
\end{align}
\textit{then}
$$\lim_{n\to\infty}\frac{1}{n\rho}\log \mathbb{E}_{P_n}[2^{\rho L(C_n(X^n))}] = H_{\alpha}(P).$$

\begin{IEEEproof}
Choose $\psi_n(x^n) = 2^{L(C_n(x^n))}$. Then, from \eqref{eq:optimal_code_length_campbell} we have
\begin{equation*}
\frac{Z_{P_n,\alpha}}{P_n(x^n)^{\alpha}} \le \psi_n(x^n) < 2\cdot \frac{Z_{P_n,\alpha}}{P_n(x^n)^\alpha}\cdot
\end{equation*}

The result follows by applying \cref{thm:bound2_unified_mismatch} and \cref{thm:bound3_unified_mismatch} with $c_n = 2$, $a_n = 1$ and $Q=P$.
\end{IEEEproof}

\medskip
\noindent
\textbf{Mismatch Case}:

Redundancy in the mismatched case of the Campbell's problem was studied in \cite{BlumerMcElice,Sundaresan}. Sundaresan showed that the difference in the normalized cumulant from the minimum when encoding according to an arbitrary uniquely decodable code is measured by the $I_{\alpha}$-divergence up-to a factor of 1 \cite{Sundaresan}. We generalize this idea as follows.

\begin{proposition} \label{prop:campbell_coding_mismatch}
Let $X$ be a random variable that assumes values from set $\mathcal{X}$ according to a probability distribution $P$. Let $\rho\in (-1,0)\cup (0,\infty)$ and $L: \mathcal{X} \to \mathbb{Z}_{+}$ be an arbitrary length function that satisfies \eqref{eq:KraftMcmilan}. Define
\begin{align}
\label{eq:difference_cumlant}
    R_c(P,L,\rho) := \frac{1}{\rho}\log \mathbb{E}_{P}[2^{\rho L(X)}] -\min_K \frac{1}{\rho}\log \mathbb{E}_{P}[2^{\rho K(X)}],
\end{align}
where 
the minimum is over all length functions $K$ satisfying \eqref{eq:KraftMcmilan}. Then, there exists a probability distribution $Q_L$ such that
\begin{align}
    \label{eq:bounds_difference_cumulant}
    I_{\alpha}(P,Q_L) - \log \eta - 1 \le R_c(P,L,\rho) \le I_{\alpha}(P,Q_L) - \log \eta,
\end{align}
where $\eta  = \sum_x 2^{-L(x)}$. 
\end{proposition}

\begin{IEEEproof}
Since $K$ satisfies \eqref{eq:KraftMcmilan}, an application of \cref{prop:lower_bound_unified} with $\psi(x) = 2^{K(x)}$ gives us $\frac{1}{\rho} \log \mathbb{E}_{P}[2^{\rho K(X)}]  \geq H_{\alpha}(P)$. Since $K(x) = \lceil \log ({Z_{P,\alpha}}/{P(x)^\alpha)\rceil}$ satisfies \eqref{eq:KraftMcmilan} and $\psi(x) = 2^{K(x)} < 2\cdot {Z_{P,\alpha}}/{P(x)^\alpha}$, applying \cref{thm:bound2_unified_mismatch} with $n=1, c_1 = 2$, and $Q = P$, we have 
\[
\frac{1}{\rho} \log \mathbb{E}_{P}[2^{\rho K(X)}]  \leq H_{\alpha}(P) + 1,
\]
that is, the minimum in \eqref{eq:difference_cumlant} is between $H_\alpha(P)$ and $H_\alpha(P) + 1$. Hence, 
\begin{align}
\frac{1}{\rho} \log \mathbb{E}_P[2^{\rho L(X)}] - H_{\alpha}(P) -1 \le R_c(P,L,\rho) \le \frac{1}{\rho}  \log  \mathbb{E}_P[2^{\rho L(X)}] - H_\alpha(P).
\label{eqn:upper_lower_bound_difference_cumulant}
\end{align}
Let us now define a probability distribution $Q_L$ as 
\begin{equation*}
Q_L(x) = \frac{2^{-L(x)/\alpha}}{\sum\nolimits_{x'}2^{-L(x')/\alpha}}\cdot
\end{equation*}
Then
\begin{equation*}
2^{L(x)} = Q_L(x)^{-\alpha} Z_{Q_L,\alpha}\cdot \frac{1}{\sum\nolimits_{x'}2^{-L(x')}} = Q_L(x)^{-\alpha} Z_{Q_L,\alpha}\cdot \frac{1}{\eta},
\end{equation*}
where $\eta = \sum_{x'}2^{-L(x')}$. Applying Propositions~\ref{thm:bound2_unified_mismatch} and \ref{thm:bound3_unified_mismatch} with $n=1$, $\psi_1(x) = 2^{L(x)}$, $a_1 = c_1 = {1}/{\eta}$, we get
\begin{equation}
\label{eqn:upper_lower_bound_mismatch_campbell}
\frac{1}{\rho} \log \mathbb{E}_{P}[2^{\rho L(X)}] = H_\alpha(P) + I_\alpha(P,Q_L) - \log \eta.
\end{equation}
Combining (\ref{eqn:upper_lower_bound_difference_cumulant}) and (\ref{eqn:upper_lower_bound_mismatch_campbell}), we get the desired result.
\end{IEEEproof}
\medskip

We note that the bound in \eqref{eq:bounds_difference_cumulant} can be loose when $\eta$ is small. For example, for a source with two symbols, say $x$ and $y$, with code lengths $L(x) = L(y) = 100$, we have $R_c(P,L,\rho) \geq I_{\alpha}(P,Q_L) + 98$. However, if one imposes the constraint $\nicefrac{1}{2} \leq \eta \leq 1$, then \eqref{eq:bounds_difference_cumulant} simplifies to
$$
|R_c(P,L,\rho) - I_{\alpha}(P,Q_L)| \le 1,
$$
which is {\rm \cite[Th.~8]{Sundaresan}}. $I_{\alpha}(P,Q_L)$ is, in a sense, the penalty when $Q_L$ does not match the true distribution $P$. In view of this, a result analogous to  \cref{prop:campbell_coding_mismatch} also holds for the Shannon source coding problem.

\subsection{Ar{\i}kan's Guessing Problem}

Let $\mathcal{X}$ be a set of objects with $|\mathcal{X}| = m$. Bob thinks of an object $X$ (a random variable) from $\mathcal{X}$ according to a probability distribution $P$. Alice guesses it by asking questions of the form ``is $X=x$?". The objective is to minimize average number of guesses required to correctly guess $X$. By a guessing strategy (or guessing function), we mean a one-one map $G:\mathcal{X}\to\{1,\ldots,m\}$, where $G(x)$ is to be interpreted as the number of questions required to guess $x$ correctly. Ar{\i}kan studied the $\rho^{\textrm{th}}$ moment of number of guesses and found upper and lower bounds in terms of R\'enyi entropy.

\medskip
\noindent
\textbf{Theorem 1 of \cite{Arikan}}:
\textit{Let $G$ be any guessing function. Then, for $\rho \in (-1,0) \cup (0,\infty)$,}
 $$\frac{1}{\rho}\log\mathbb{E}_{P}\big[G(X)^\rho\big]\geq H_\alpha(P) - \log (1+\ln m).$$
\begin{IEEEproof}
Let $G$ be any guessing function. Let $\psi(x) = G(x)$. Then, we have $\sum_{x \in \mathcal{X}} \psi(x)^{-1} = \sum_{x \in \mathcal{X}} {1}/{G(x)} = \sum^m_{i=1} {1}/{i} \le 1 + \ln m$. An application of \cref{prop:lower_bound_unified} with $k = 1 + \ln m$ yields the desired result.
\end{IEEEproof}

Ar{\i}kan showed that an optimal guessing function guesses according to the decreasing order of $P$-probabilities with ties broken using an arbitrary but fixed rule \cite{Arikan}. He also showed that normalized cumulant of an optimal guessing function is bounded above by the R\'enyi entropy. Next, we present a proof of this using our general framework.

\medskip
\noindent
\textbf{Proposition 4 of \cite{Arikan}}:
\textit{If $G^*$ is an optimal guessing function, then for $\rho \in (-1,0) \cup (0,\infty)$,}
$$ \frac{1}{\rho}\log\mathbb{E}_{P}\big[G^*(X)^\rho\big]\leq H_\alpha (P).$$
\begin{IEEEproof}
Let us rearrange the probabilities $\{P(x) , x \in \mathcal{X}\}$ in non-increasing order, say
$$p_1 \geq p_2 \geq \cdots \geq p_m.$$
Then, the optimal guessing function $G^{*}$ is given by $G^{*}(x) = i$ if $P(x) = p_i$. Let us index the elements in set $\mathcal{X}$ as $\{x_1, x_2, \ldots, x_m\}$, according to the decreasing order of their probabilities. Then, for $i\in \{1,\dots, m\}$, we have
\begin{equation}
\label{eqn:optimal_guessing_eqn}
    \frac{Z_{P,\alpha}}{P(x_i)^{\alpha}} = \frac{\sum^m_{j=1} p^{\alpha}_j}{p^{\alpha}_i} \geq i =  G^{*}(x_i).
\end{equation}
That is, $G^{*}(x) \leq \frac{Z_{P,\alpha}}{P(x)^{\alpha}}$ for $x \in \mathcal{X}$. Now, an application of \cref{thm:bound2_unified_mismatch} with $n=1$, $Q = P$, $\psi_1(x) = G^{*}(x)$ and $c_1 = 1$, gives us
$$
\frac{1}{\rho} \log \mathbb{E}_{P}[G^{*}(X)^{\rho}] =  \frac{1}{\rho} \log \mathbb{E}_{P}[\psi(X)^{\rho}] \leq H_{\alpha}(P) + I_{\alpha}(P,P) + \log 1 = H_{\alpha}(P).
$$
\end{IEEEproof}

Ar{\i}kan also proved that the upper bound of R\'enyi entropy can be achieved by guessing long sequences of letters in an i.i.d. fashion. 

\medskip
\noindent
\textbf{Proposition 5 of \cite{Arikan}}:
\textit{Let $X_1,X_2,\ldots,X_n$ be a sequence of i.i.d. guesses. Let $G_n^*(X_1,\ldots,X_n)$ be an optimal guessing function. Then, for  $\rho \in (-1,0) \cup (0,\infty)$,}
$$ \lim_{n \to \infty}\frac{1}{n\rho} \log \mathbb{E}_{P_n}[G_n^*(X_1,X_2,\ldots,X_n)^\rho] = H_\alpha(P).$$
\begin{IEEEproof}
Let $G_n^*$ be the optimal guessing function from $\mathcal{X}^n$ to $\{1,2,\ldots, m^n\}$. An application of Corollary \ref{corr:sequence_bound_lower} with $\psi_n(x^n) = G_n^*(x^n)$ and $k_n = 1 + n\ln m$ yields
\begin{eqnarray}
\label{eq:massey_thm3_eq1}
\liminf_{n \to \infty} \frac{1}{n\rho} \log \mathbb{E}_{P_n}[G_n^*(X^n)^{\rho}] & \geq & H_{\alpha}(P) - \limsup_{n \to \infty} \frac{\log (1 + n\ln m)}{n}\nonumber\\
& = & H_{\alpha}(P).
\end{eqnarray}
As in the proof of the previous result, we know that $G^{*}(x^n) \leq \frac{Z_{P_n,\alpha}}{P_n(x^n)^{\alpha}}$ for $x^n \in \mathcal{X}^n$. Hence an application of \cref{thm:bound2_unified_mismatch} with $Q_n = P_n$, $\psi_n(x^n) = G_n^{*}(x^n)$, and $c_n = 1$ yields
\begin{align} \label{eq:massey_thm3_eq2}
\limsup_{n \to \infty} \frac{1}{n\rho} \log \mathbb{E}_{P_n}[G_n^{*}(X^n)^{\rho}] \leq H_{\alpha}(P).
\end{align}
Combining \eqref{eq:massey_thm3_eq1} and \eqref{eq:massey_thm3_eq2} we get the desired result.
\end{IEEEproof}
Henceforth we shall denote the optimal guessing function corresponding to a probability distribution $P$ by $G_P$.

\medskip
\noindent
\textbf{Mismatch Case}:

Suppose Alice does not know the true underlying probability distribution $P$, and  guesses according to some guessing function $G$. The following proposition tells us that the penalty for deviating from the optimal guessing function can be measured by $I_{\alpha}$-divergence.

\begin{proposition}
\label{prop:guessing_mismatch_lowerbound}
\textit{Let $G$ be an arbitrary guessing function. Then, for $\rho \in (-1,0) \cup (0,\infty)$, there exists a probability distribution $Q_G$ on $\mathcal{X}$ such that}
$$\frac{1}{\rho}\log\mathbb{E}_{P}[G(X)^{\rho}] \ge H_\alpha (P) + I_\alpha (P, Q_G)- \log (1 + \ln m).$$
\end{proposition}
 
\begin{IEEEproof}
Let $G$ be a guessing function. Define a probability distribution $Q_G$ on $\mathcal{X}$ as
\begin{equation}
\label{eqn:defn-Q_G}
 Q_G(x) = \frac{G(x)^{-1/\alpha}}{\sum\nolimits_{x' \in \mathcal{X}} G(x')^{-1/\alpha}}.
\end{equation}
Then, we have
$$
\frac{Z_{Q_G,\alpha}}{Q_G(x)^{\alpha}} = G(x) \sum \nolimits_{x' \in \mathcal{X}} \frac{1}{G(x')} \le G(x)\cdot (1 + \ln m)\cdot
$$
 Now, an application of \cref{thm:bound3_unified_mismatch} with $n=1$, $\psi_1(x) = G(x)$, and $a_1 = {1}/{(1 + \ln m)}$  yields the desired result.
\end{IEEEproof}

\noindent
A converse result is the following.
\medskip

\noindent
\textbf{Proposition 1 of \cite{Sundaresan}}: \textit{Let $G_Q$ be an optimal guessing function associated with $Q$. Then, for $\rho \in (-1,0) \cup (0,\infty)$,}
\begin{align*}
 \frac{1}{\rho}\log \mathbb{E}_{P}\big[G_Q(X)^\rho \big]	\leq H_\alpha(P) + I_\alpha(P,Q),
\end{align*}
\textit{where the expectation is with respect to $P$.}
\begin{IEEEproof}
Let us rearrange the probabilities $(Q(x) , x \in \mathcal{X})$ in non-increasing order, say
$$q_1 \geq q_2 \geq \cdots \geq q_m.$$

By definition, $G_Q(x) = i$ if $Q(x) = q_i$. Then, as in \eqref{eqn:optimal_guessing_eqn}, we have $G_Q(x) \leq {Z_{Q,\alpha}}/{Q(x)^{\alpha}}  \text{ for } x \in \mathcal{X}$. Hence an application of \cref{thm:bound2_unified_mismatch} with $n=1$, $\psi_1(x) = G_Q(x)$, and $c_1 = 1$ proves the result.
\end{IEEEproof}

Observe that, given a guessing function $G$, if we apply the above proposition for $Q = Q_G$, where $Q_G$ is as in \eqref{eqn:defn-Q_G}, then we get
\begin{align*}
 \frac{1}{\rho}\log \mathbb{E}_{P_n}\big[G_{Q_G}(X)^\rho \big]	\leq H_\alpha(P) + I_\alpha(P,Q_G).
\end{align*}

Thus, the above two propositions can be combined to state the following, which is analogous to \cref{prop:campbell_coding_mismatch} (refer \cref{sec:campbell_coding}).

\medskip
\noindent
\textbf{Theorem 6 of \cite{Sundaresan}}:
\textit{Let $G$ be an arbitrary guessing function and $G_P$ be the optimal guessing function for $P$. For $\rho \in (-1,0) \cup (0,\infty)$, let}
\begin{equation*}
 R_{g}(P,G,\rho) := \frac{1}{\rho}\log \mathbb{E}_{P}\big[G(X)^\rho \big] - \frac{1}{\rho}\log \mathbb{E}_{P}\big[G_P(X)^\rho \big]. 
\end{equation*}
\textit{Then, there exists a probability distribution $Q_G$ such that}
\begin{equation*}
    | R_{g}(P,G,\rho) - I_\alpha (P, Q_G)| \le \log (1 + \ln m).
\end{equation*}

\subsection{Memoryless Guessing}
\label{subsec:memoryless}

In memoryless guessing, the setup is similar to that of Ar{\i}kan's guessing problem except that this time the  guesser Alice comes up with guesses independent of her previous guesses. Let $\hat{X}_1, \hat{X}_2, \ldots$ be Alice's sequence of independent guesses according to a distribution $\widehat{P}$. The guessing function in this problem is defined as,
\begin{align*}
G_{\widehat{P}}(X) := \inf\{i \geq 1 : \hat{X}_{i} = X\},
\end{align*}
that is, the number of guesses until a successful guess. Sundaresan \cite{HanS19}, inspired by Ar{\i}kan's result, showed that the minimum expected number of guesses required is $\exp\{H_{\frac{1}{2}}(P)\}$, and the distribution that achieves this, is surprisingly not the underlying distribution $P$, but the ``tilted distribution" $\widehat{P}^*(x) := {\sqrt{P(x)}}/{\sum_y \sqrt{P(y)}}$.

Unlike in Arikan's guessing problem, Huleihel et al. \cite{Huleihel} minimized what are called {\em factorial moments}, defined for $\rho \in \mathbb{Z}_{+}$ as,
\begin{align*}
 V_{\widehat{P}, \, \rho} (X)=\frac{1}{\rho !} \prod \nolimits_{l=0}^{\rho -1}
\big (G_{\widehat{P}}(X) + l\big ).
\end{align*}
Huleihel et al. \cite{Huleihel} studied the following problem.
\begin{align*}
\text{Minimize} \quad\mathbb{E}_{P}\left[V_{\widehat{P}, \,  \rho} (X)\right],
\end{align*}
over all $\widehat{P}\in \mathcal{P}$, where $\mathcal{P}$ is the probability simplex, that is, $\mathcal{P} = \{(P(x))_{x \in \mathcal{X}} : P(x)\geq 0,\sum_x P(x)=1\}$. Let $\widehat{P}^{*}$ be the optimal solution of the above problem. 

\medskip
\noindent
\textbf{Theorem 1 of \cite{Huleihel}}: \textit{For any integer $\rho>0$, we have}
$$ \frac{1}{\rho} \log \mathbb{E}_{P}\left[V_{\widehat{P}^{*}, \,  \rho} (X) \right]
 = H_{\alpha}(P)$$
\textit{and $\widehat{P}^*(x)={P(x)^\alpha}/{Z_{P,\alpha}}.$}
\begin{IEEEproof}
From \cite{Huleihel}, we know that
\begin{align}
\mathbb{E}_{P}\left[V_{\widehat{P}, \,  \rho} (X)\right] = \mathbb{E}_{P}[\widehat{P}(X)^{-\rho}]. \label{eqn:factorial_moment}
\end{align}

Now, the result follows from \cref{prop:lower_bound_unified} with $\psi(x) = \widehat{P}(x)^{-1} $ and $k = 1$. Indeed, since $\widehat{P}$ is a probability distribution, we have $\sum_{x \in \mathcal{X}} \psi(x)^{-1} = \sum_{x \in \mathcal{X}} \widehat{P}(x) = 1$. Hence $\frac{1}{\rho} \log \mathbb{E}_{P} \left[V_{\widehat{P},\,\rho}(X)\right] \geq H_{\alpha}(P)$, and the lower bound is attained by $\widehat{P}^*(x)={P(x)^\alpha}/{Z_{P,\alpha}}$.
\end{IEEEproof}

For sequence of random guesses, above theorem can be stated in the following way. Let $\hat{X}^n = (\hat{X}_1,\ldots,\hat{X}_n$), where $X_i$'s are i.i.d. guesses, drawn from $\mathcal{X}^n$ with distribution $\widehat{P}_n$ --- the $n$-fold product distribution of $\widehat{P}$ on $\mathcal{X}^n$. If the true underlying distribution is $P_n$,  then
\begin{align*}
\lim_{n\to\infty}\frac{1}{n\rho} \log \mathbb{E}_{P_n}\left[V_{\widehat{P}^{*}_n \, \rho} (X^n) \right] = H_{\alpha}(P), 
\end{align*}
where $\widehat{P}_n^*(x)={P_n(x)^\alpha}/{Z_{P_n,\alpha}}$. For the mismatched case, we have the following result.

\begin{proposition} \label{prop:memoryless_mismatch}
If the true underlying probability distribution is $P$ but Alice assumes it as $Q$ and guesses according to its optimal one, namely $\widehat{Q}^{*}(x) = Q(x)^{\alpha} / Z_{Q,\alpha}$, then
$$\frac{1}{\rho} \log \mathbb{E}_{P}\left[V_{\widehat{Q}^*, \, \rho}(X)\right] = H_{\alpha}(P) + I_{\alpha}(P,Q).$$
\end{proposition}
\begin{IEEEproof}
Due to \eqref{eqn:factorial_moment}, the result follows easily by taking $n=1$, $\psi_1(x) = \widehat{Q}^*(x)^{-1}$, $c_1=1, a_1=1$ in Propositions \ref{thm:bound2_unified_mismatch} and \ref{thm:bound3_unified_mismatch}.
\end{IEEEproof}

\subsection{Tasks Partitioning Problem}

\emph{Encoding of Tasks} problem studied by Bunte and Lapidoth \cite{Bunte} can be phrased the following way. Let $\mathcal{X}$ be a finite set of tasks. A task $X$ is randomly drawn from $\mathcal{X}$ according to a probability distribution $P$, which may correspond to the frequency of occurrences of tasks. Suppose these tasks are associated with $M$ keys. Typically, $M < |\mathcal{X}|$. Due to the limited availability of keys, more than one task may be associated with a single key. When a task needs to be performed, the key associated with it is pressed. Consequently, all tasks associated with this key will be performed. The objective in this problem is to minimize the number of redundant tasks performed. Usual coding techniques suggest assigning tasks with high probability to individual keys and leaving the low probability tasks unassigned. But for an individual, all tasks can be equally important. It may just be the case that some tasks may have a higher frequency of occurrence than others. If $M \geq |\mathcal{X}|$, then one can perform tasks without any redundancy. However, Bunte and Lapidoth \cite{Bunte} showed that, even when $M<|\mathcal{X}|$, one can accomplish the tasks with much less redundancy on average, provided the underlying probability distribution is different from the uniform distribution. 

Let $\mathcal{A} = \{\mathcal{A}_1,\mathcal{A}_2,\ldots,\mathcal{A}_M\}$ be a partition of $\mathcal{X}$ that corresponds to the assignment of tasks to $M$ keys. Let $A(x)$ be the cardinality of the subset containing $x$ in the partition. We shall call $A$ the {\em partition function} associated with partition $\mathcal{A}$.

\medskip
\noindent
\textbf{Theorem I.1 of \cite{Bunte}}: The following results hold.
\begin{enumerate}
    \item[(a)]  Let $\rho \in (-1,0) \cup (0, \infty)$. For any partition of $\mathcal{X}$ of size $M$ with partition function $A$, we have
		$$\frac{1}{\rho}\log \mathbb{E}_{P}[A(X)^\rho]\geq H_\alpha(P)-\log M.$$
		
	\item[(b)] Let $\rho \in (0, \infty)$. If $M>\log |\mathcal{X}|+2$, then there exists a partition of $\mathcal{X}$ of size at most $M$ with partition function ${A}$ such that
		$$
		1 \leq \mathbb{E}_{P}[\widehat{A}(X)^\rho] \leq 1+2^{\rho(H_\alpha(P)-\log\tilde{M})},
		$$
	where 
	\begin{align}
	    \tilde{M} := {(M-\log |\mathcal{X}| - 2)}/{4}. \label{eq:m_tilde}
	\end{align}
	\end{enumerate}

\noindent
\begin{IEEEproof}
\textit{Part (a)}: Let $\psi(x) = A(x)$. Then, we have
$\sum_{x \in \mathcal{X}} \psi(x)^{-1} = \sum_{x \in \mathcal{X}} A(x)^{-1} = M$ \cite[Prop.~III-1]{Bunte}. Now, an application of \cref{prop:lower_bound_unified} with $k = M$ give us the desired result.
\medskip

\noindent
\textit{Part (b)}: For the proof of this part, we refer to \cite{Bunte}.
\end{IEEEproof}

Bunte and Lapidoth also proved the following limit results.

\medskip
\noindent
\textbf{Theorem I.2  of \cite{Bunte}}: Let  $\rho  >0$. Then for every $n\geq 1$ there exists a partition $\mathcal{A}_n$ of $\mathcal{X}^n$ of size at most $M^n$ with associated partition function $A_n$  such that 
$$\lim_{n\to \infty} \mathbb{E}_{P_n} [A_n(X^n)^{\rho}] = \begin{cases}
1 & \textrm{if } \log M > H_{\alpha}(P) \\ 
\infty & \textrm{if } \log M < H_{\alpha}(P),
\end{cases}
$$
where $X^n:=(X_1,\dots,X_n)$.

It should be noted that, in a general set-up of the tasks partitioning problem, it is not necessary that the partition size is of the form $M^n$; it can be some $M_n$ (a function of $n$). Consequently, we have the following result.
\begin{proposition} \label{thm:tasks_partition_refined}
Let $\{M_n\}$ be a sequence of positive integers such that $M_n \geq n \log |\mathcal{X}| + 3$,  and 
$$\gamma:= \lim_{n\to\infty} \frac{\log M_n}{n}$$ exists. Then, for $\rho > 0$, there exists a sequence of partitions of $\mathcal{X}^n$ of size at most $M_n$ with partition functions $A_n$ such that
\begin{enumerate}
	\item[(a)]	$$\lim_{n\to \infty} \mathbb{E}_{P_n} [A_n(X^n)^{\rho}] = 1 \qquad \text{if } \gamma > H_{\alpha}(P),$$
		
		\item[(b)]$$\lim_{n\to\infty}\frac{1}{n\rho}\log \mathbb{E}_{P_n}[A_n(X^n)^{\rho}] = H_\alpha(P) - \gamma \qquad \text{if } \gamma < H_{\alpha}(P).$$
\end{enumerate}

\end{proposition}
\begin{IEEEproof}
Let 
\begin{align}
\tilde{M}_n := (M_n - n \log |\mathcal{X}|  - 2)/4. \label{eq:mn_tilde}    
\end{align}
We first claim that $\lim_{n \to \infty} \frac{\log \tilde{M}_n}{n} = \gamma$. Indeed, since $\frac{\log(1/4)}{n} \leq \frac{\log \tilde{M}_n}{n} < \frac{\log M_n}{n}$, when $\gamma = 0$, we have $\lim_{n \to \infty} \frac{\log \tilde{M}_n}{n} = 0$. On the other hand, when $\gamma > 0$, we can find an $n_{\gamma}$ such that $M_n \geq 2^{\gamma n /2} \, \forall n \geq n_{\gamma}$. Thus, we have $\lim_{n \to \infty} \frac{n}{M_n} = 0$. Consequently,  
$$\lim_{n \to \infty} \frac{\log \tilde{M}_n}{n} = \lim_{n \to \infty} \frac{\log {M}_n}{n} + \lim_{n \to \infty} \frac{1}{n} \log \left(1 - \frac{(n \log |\mathcal{X}| +2 )}{M_n}  \right) - \lim_{n \to \infty} \frac{2}{n} = \gamma. $$
\noindent
This proves the claim. From Theorem I.1 of [11], for any $n \geq 1$ and $M_n > n \log |\mathcal{X}| + 2$, there exists a partition $\mathcal{A}_n$ of $\mathcal{X}^n$ of size at most $M_n$ such that the associated partition function $A_n$ satisfies
\begin{align*}
\mathbb{E}_{P_n}[A_n(X^n)^{\rho}] & \leq  1 + 2^{\rho(H_{\alpha}(P_n)- \log \tilde{M}_n)}\\
& =  1 + 2^{n \rho\big(H_{\alpha}(P)- \frac{\log \tilde{M}_n}{n}\big)}.
\end{align*}

\noindent
\textit{Part (a)}:  When $\gamma > H_{\alpha}(p)$, let us choose $\epsilon  = (\gamma - H_{\alpha}(P))/2 > 0$. Then, there exists an $n_{\epsilon}$ such that $\frac{\log \tilde{M}_n}{n} \geq \gamma - \epsilon \, \forall n \geq n_{\epsilon}$. Thus, we have 
\begin{align*}
 \mathbb{E}_{P_n}[A_n(X^n)^{\rho}]  \leq  1 +  2^{n \rho (H_{\alpha}(P) - \gamma + \epsilon )} = 1 +  2^{-n \rho (\gamma - H_{\alpha}(P))/2} \quad \forall n \geq n_{\epsilon}
\end{align*}
Consequently, $\limsup_{n \to \infty} \mathbb{E}_{P_n}[A_n(X^n)^{\rho}] \leq 1$. We also note that $A_n(x^n) \geq 1$ for all $x^n \in \mathcal{X}^n$. 

\noindent
Thus, $\liminf_{n \to \infty} \mathbb{E}_{P_n}[A_n(X^n)^{\rho}] \geq 1$.

\noindent
\textit{Part (b)}: For any $\epsilon  > 0$, there exists an $n_{\epsilon}$ such that $\frac{\log \tilde{M}_n}{n} \geq \gamma - \epsilon \, \forall n \geq n_{\epsilon}$. Thus, we have 
\begin{align*}
\mathbb{E}_{P_n}[A_n(X^n)^{\rho}] & \leq  1 + 2^{n \rho(H_{\alpha}(P)- \gamma + \epsilon )} \leq 2^{1 + n \rho(H_{\alpha}(P)- \gamma + \epsilon )} \quad \forall n \geq n_{\epsilon}
\end{align*}

\noindent
Hence, we have  
$$\limsup_{n \to \infty} \frac{1}{n\rho} \log \mathbb{E}_{P_n}[A_n(X^n)^{\rho}] \leq H_{\alpha}(P)  - \gamma +\epsilon \quad \forall \epsilon > 0$$

\noindent
Further, an invocation of Corollary~\ref{corr:sequence_bound_lower} with $\psi_n(x^n) = A_n(x^n) $ and $k_n = \sum_{x^n \in \mathcal{X}^n} 1/A_n(x^n) = M_n$ gives us
$$\liminf_{n \to \infty} \frac{1}{n\rho} \log \mathbb{E}_{P_n}[A_n(X^n)^{\rho}] \geq H_{\alpha}(P) - \limsup_{n \to \infty} \frac{\log M_n}{n} = H_{\alpha}(P) - \gamma.$$
\end{IEEEproof}

\begin{remark}
It is interesting to note that when $\gamma <  H_{\alpha}(P)$, in addition to the fact that $\lim_{n \to \infty} \mathbb{E}_{P_n}[A_n(X^n)^{\rho}] = \infty$, we also have  $\mathbb{E}_{P_n}[A_n(X^n)^{\rho}] \approx 2^{n\rho (H_{\alpha}(P) - \gamma)}$ for large values of $n$.
\end{remark}

\medskip
\noindent
\textbf{Mismatch Case}:

\noindent
 Let us now suppose that one does not know the true  underlying  probability  distribution $P$, but arbitrarily partitions $\mathcal{X}$. Then, the penalty due to such a partition can be measured by the $I_{\alpha}$-divergence as stated in the following theorem.

\begin{proposition}
 \label{prop:task_partition_mismatch_lb}
Let $\rho \in (-1,0) \cup (0,\infty)$. Let $\mathcal{A}$ be a partition of $\mathcal{X}$ of size $M$ with partition function $A$. Then, there exists a probability distribution $Q_{A}$ on $\mathcal{X}$ such that 
$$\frac{1}{\rho}\log \mathbb{E}_{P}[A(X)^\rho] = H_\alpha(P) + I_{\alpha}(P,Q_A) -\log M.$$
\end{proposition}

\begin{IEEEproof}
Define a probability distribution 
$Q_A = \{Q_A(x),~x \in \mathcal{X}\}$ as
$$Q_A(x) := \frac{A(x)^{-1/\alpha}}{\sum_{x' \in \mathcal{X}} A(x')^{-1/\alpha}}\cdot$$
Then
$$\frac{Z_{Q_A,\alpha}}{Q_A(x)^{\alpha}} = A(x) \sum \nolimits_{x' \in \mathcal{X}} \frac{1}{A(x')} = A(x)\cdot M,$$
where the last equality follows due to \cite[Prop.~III.1]{Bunte}. Rearranging terms, we have $A(x)=\frac{Z_{Q_A,\alpha}}{M \cdot Q_A(x)^{\alpha}}$. Hence an application of Propositions \ref{thm:bound2_unified_mismatch} and \ref{thm:bound3_unified_mismatch} with $n=1$, $\psi_1(x) = A(x)$, $c_1 = 1/M$, $a_1 = 1/M$, and $Q=Q_A$ yields the desired result.
\end{IEEEproof}

\noindent
A converse result is the following.

\begin{proposition} \label{prop:task_partition_mismatch_ub}
Let $X$ be a random task from $\mathcal{X}$ following distribution
$P$ and $\rho\in (0,\infty)$. Let $Q$ be  another  distribution on $\mathcal{X}$. If $M > \log |\mathcal{X}| +2$, then there exists a partition $\mathcal{A}_Q$ (with an associated partition function $A_Q$) of $\mathcal{X}$ of size at most $M$ such that
$$ \mathbb{E}_{P}[A_Q(X)^\rho] \leq 1+2^{\rho(H_\alpha(P) + I_\alpha(P,Q)-\log\tilde{M})},$$
where $\tilde{M}$ is as in \eqref{eq:m_tilde}.
\end{proposition}

\begin{IEEEproof}
Similar to proof of Theorem I.1 of \cite{Bunte}.
\end{IEEEproof}

\section{Ordered Tasks Partitioning Problem}
\label{sec:generalized_task}

In Bunte and Lapidoth's tasks partitioning problem \cite{Bunte}, one is interested in the average number of tasks associated with a key. However, in some scenarios, it might be more important to minimize the average number of redundant tasks performed, before the intended task. To achieve this, tasks  associated  with  a  key  should  be  performed  in  decreasing  order  of  their probabilities.  With such a strategy in place, this problem draws parallel  with Ar{\i}kan's guessing problem \cite{Arikan}.

Let $\mathcal{A} = \{\mathcal{A}_1,\mathcal{A}_2,\ldots,\mathcal{A}_M\}$ be a partition of $\mathcal{X}$ that corresponds to the assignment of tasks to $M$ keys. Let $N(x)$ be the number of redundant tasks performed until and including the intended task $x$. We refer to $N(\cdot)$ as the {\em count function} associated with partition $\mathcal{A}$. We suppress the dependence of $N$ on $\mathcal{A}$ for the sake of notational convenience. If $X$ denotes the intended task, then we are interested in the $\rho^{\textrm{th}}$ moment of number of tasks performed, that is,  $\mathbb{E}_{P}[N(X)^\rho]$, where $\rho\in (-1,0)\cup (0,\infty)$. 

\begin{lemma}
\label{lem:Kraft_gen_partition}
For any count function associated with a partition of size $M$, we have
\begin{equation}
\label{eqn:Kraft_gen_partition}
\sum_{x\in\mathcal{X}}\frac{1}{N(x)} \le M \left[1 + \ln \left ( \frac{|\mathcal{X}|}{M} \right)\right].
\end{equation}
\end{lemma}
\begin{IEEEproof}
For a partition $ \mathcal{A}=\{A_1,A_2,\dots,A_M \}$ of $\mathcal{X}$, observe that
\begin{equation}
\label{eqn:sum_count_function}
\sum_{x\in\mathcal{X}}\frac{1}{N(x)} = \left (1+\frac{1}{2}+\dots+\frac{1}{|A_1|} \right ) + \dots + \left(  1+\frac{1}{2}+\dots+\frac{1}{|A_M|}\right).
\end{equation}
Since
$$1+\frac{1}{2}+\dots+\frac{1}{|A_k|} \le 1 + \ln |A_k|,$$
for any $k\in\{1,\dots,M\}$, we have 
\begin{align}
     \sum_{x\in\mathcal{X}}\frac{1}{N(x)} & \le M + \ln(|A_1|\dots |A_M|) \nonumber \\
     & = M[1 + \ln(|A_1|\dots |A_M|)^{1/M}] \nonumber \\
     & \overset{(a)}{\leq} M\left[1 + \ln\left(\frac{|A_1| + \dots + |A_M|}{M}\right)\right] \nonumber \\
     & =  M \left[1 + \ln\left ( \frac{|\mathcal{X}|}{M} \right)\right], \label{eq:ord_tasks_bound}
\end{align}
where (a) follows due to the AM-GM inequality.
\end{IEEEproof}

\begin{proposition} \label{thm:gen_task_partition_thm1}
Let $\rho \in (-1,0) \cup (0, \infty)$ and $X$ be a random task from $\mathcal{X}$ following distribution $P$. Then the following hold.
	\begin{enumerate}
		\item[(a)] For any partition of $\mathcal{X}$ of size $M$, we have
		\begin{eqnarray}
        \frac{1}{\rho}\log\mathbb{E}_{P}[N(X)^\rho] \ge H_\alpha(P) - \log\{M [1 + \ln \left ({|\mathcal{X}|}/{M} \right)]\} 
        \end{eqnarray}

        \item[(b)] Let $M>\log |\mathcal{X}|+2$. Then, for $\rho > 0$, then there exists a partition of $\mathcal{X}$ of size at most $M$ with count function ${N}$ such that
		$$
		1 \leq \mathbb{E}_{P}[{N}(X)^\rho] \leq 1+2^{\rho(H_\alpha(P)-\log\tilde{M})},
		$$
	where $\tilde{M}$ is as in \eqref{eq:m_tilde}.
	\end{enumerate}
\end{proposition}
\begin{IEEEproof} 
\textit{Part (a):} Applying \cref{prop:lower_bound_unified} with
$k = M \left[1 + \ln \left ({|\mathcal{X}|}/{M} \right)\right]$ and $\psi(x)= {N(x)}$,
we get the desired result.

\medskip
\noindent
\textit{Part (b):} If $A$ and $N$ are respectively the partition and count functions of a partition $\mathcal{A}$, then we have $1 \leq N(x) \leq  A(x)$ for $x\in\mathcal{X}$. Once we observe this, the proof is same as Theorem I.1 (b) of \cite{Bunte}.
\end{IEEEproof}

\begin{proposition} \label{thm:gen_task_partition_thm2}
Let $\{M_n\}$ be a sequence of positive integers such that $M_n \geq n \log |\mathcal{X}| + 3$,  and 
$\gamma:= \lim_{n\to\infty} {\log M_n}/{n}$ exists. Then, there exists a sequence of partitions of $\mathcal{X}^n$ of size at most $M_n$ with count functions $N_n$ such that 
\begin{enumerate}
		\item[(a)] $$\lim_{n\to \infty} \mathbb{E}_{P_n} [N_n(X^n)^{\rho}] = 1\qquad \text{if } \gamma > H_{\alpha}(P),$$
		\item[(b)]$$\lim_{n\to\infty}\frac{1}{n\rho}\log \mathbb{E}_{P_n}[N_n(X^n)^{\rho}] = H_\alpha(P) - \gamma \qquad \text{if } \gamma < H_{\alpha}(P).$$
\end{enumerate}
\end{proposition}
\begin{IEEEproof}
Similar to proof of \cref{thm:tasks_partition_refined}.
\end{IEEEproof}

\begin{remark}
\begin{itemize}
    \item[a).] If we choose the trivial partition, namely $\mathcal{A}_n = \{\mathcal{X}^n\}$, then the ordered tasks partitioning problem simplifies to Ar{\i}kan's guessing problem, that is, we have $M_n = 1$, $N_n(x^n)= G_n(x^n)$ and \eqref{eqn:Kraft_gen_partition} simplifies to
    \[
    \sum\limits_{x^n \in \mathcal{X}^n}\frac{1}{G_n(x^n)}\le 1+ n \ln  |\mathcal{X}|.
    \]
    Hence, all results pertaining to the Ar{\i}ken's guessing problem can be derived from the ordered tasks partitioning problem.

    \item[b)] Structurally, ordered tasks partitioning problem differs from Bunte and Lapidoth's problem only due the factor $1 + \ln(|\mathcal{X}|/M)$ in \eqref{eq:ord_tasks_bound}. While this factor matters for one-shot results, for a sequence of i.i.d. tasks, this factor vanishes asymptotically.
\end{itemize}
\end{remark}

\noindent
\textbf{Mismatch Case}:
\noindent

Let us now suppose that one does not know the true  underlying  probability  distribution $P$, but arbitrarily partitions $\mathcal{X}$ and executes tasks within each subset of this partition in an arbitrary order. Then, the penalty due to such a partition and ordering can be measured by the $I_{\alpha}$-divergence as stated in the following propositions.

\begin{proposition} \label{thm:gen_task_partition_mismatch_lb}
Let $\rho \in (-1,0) \cup (0,\infty)$. Let $\mathcal{A}$ be a partition of $\mathcal{X}$ of size $M$ with count function $N$. Then, there exists a probability distribution $Q_{N}$ on $\mathcal{X}$ such that 
$$\frac{1}{\rho}\log \mathbb{E}_{P}[N(X)^\rho] \ge H_\alpha(P) + I_{\alpha}(P,Q_N) - \log \{M \left[1 + \ln \left ({|\mathcal{X}|}/{M} \right)\right]\}.$$
\end{proposition}
\medskip
\noindent

\begin{IEEEproof}
Define a probability distribution
$Q_N = \{Q_N(x),~x \in \mathcal{X}\}$ as
$$
Q_N(x) := \frac{N(x)^{-1/\alpha}}{\sum_{x' \in \mathcal{X}} N(x')^{-1/\alpha}}\cdot
$$
Then, by Lemma \ref{lem:Kraft_gen_partition}, we have
\begin{eqnarray*}
\frac{Z_{Q_N,\alpha}}{Q_N(x)^{\alpha}} & = & N(x) \sum \nolimits_{x' \in \mathcal{X}} \frac{1}{N(x')} \le N(x)\cdot M [1 + \ln ({|\mathcal{X}|}/{M})].
\end{eqnarray*}
Now, an application of \cref{thm:bound3_unified_mismatch}  with $n=1$, $\psi_1(x) = N(x)$, $Q = Q_N$, and $a_1 = {1}/{M [1 + \ln ({|\mathcal{X}|}/{M})]}$ yields the desired result.
\end{IEEEproof}

\medskip

\noindent
A converse result is the following.

\begin{proposition} \label{thm:gen_task_partition_mismatch_ub}
Let $X$ be a random task from $\mathcal{X}$ following distribution
$P$. Let $Q$ be  another  distribution on $\mathcal{X}$. If $M > \log |\mathcal{X}| +2$, then there exists partition $\mathcal{A}_Q$ (with an associated count function $N_Q$) of $\mathcal{X}$ of size at most $M$ such that,
$$ \mathbb{E}_{P}[N_Q(X)^\rho] \leq 1 + 2^{\rho(H_\alpha(P) + I_\alpha(P,Q)-\log\tilde{M})}\qquad \text{if } \rho\in (0,\infty), $$
where $\tilde{M}$ is as in \eqref{eq:m_tilde}.
\end{proposition}

\begin{IEEEproof}
Identical to the proof of \cref{thm:gen_task_partition_thm1}(b).
\end{IEEEproof}

\section{Operational Connection Among the Problems}
\label{sec:connection}
\begin{figure}[t]
    \centering
    \includegraphics[scale=0.75]{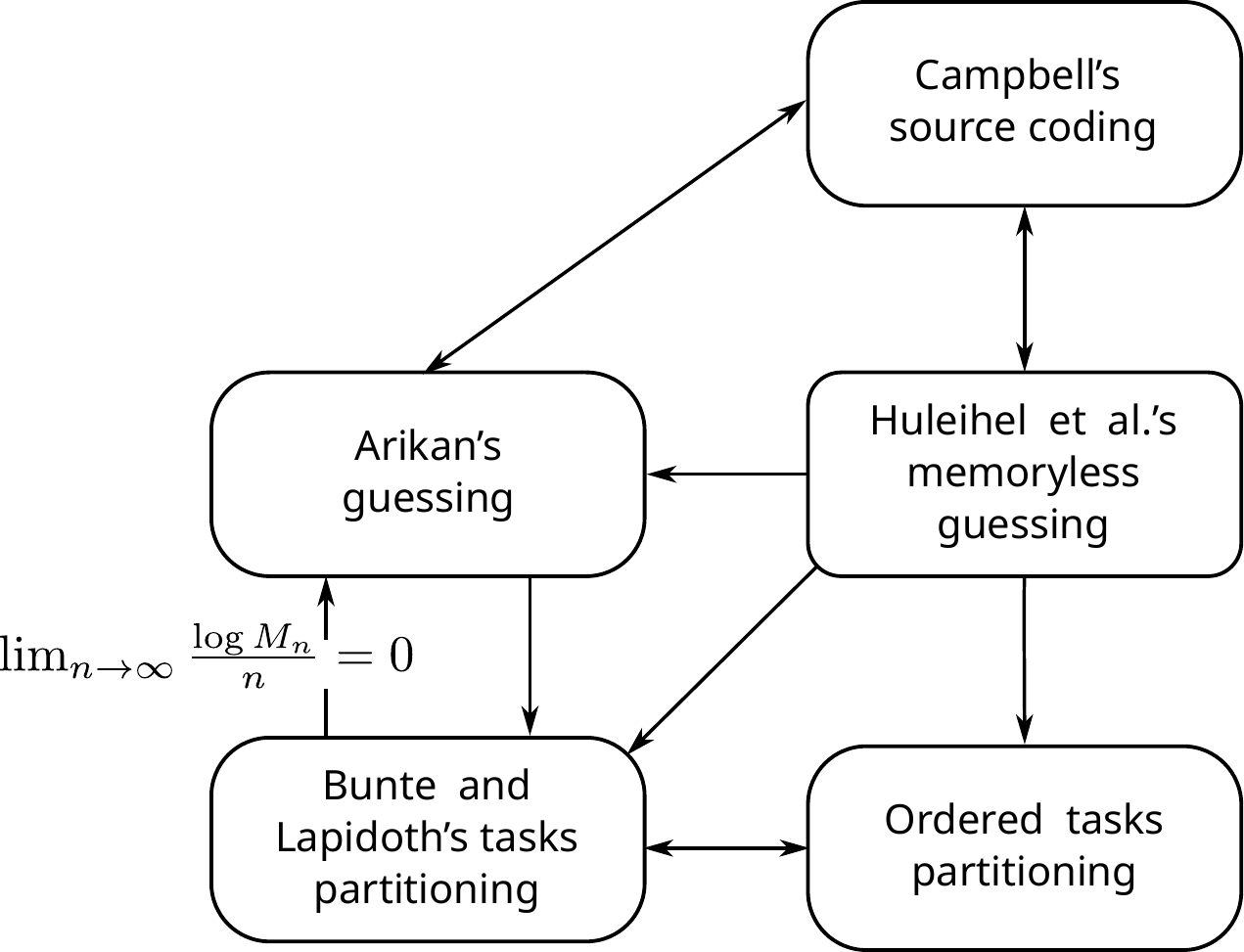}
    \caption{Relationships established among the five problems. A directed arrow from problem $A$ to problem $B$ means knowing optimal or asymptotically optimal solution of $A$ helps us find the same in $B$.}
    \label{fig:connection}
\end{figure}

In this section, we establish an operational relationship among the five problems (refer \cref{fig:connection}) that we studied in the previous section. The relationship we are interested in is ``Does knowing an optimal or asymptotically optimal solution in one problem helps us find the same in another?'' In fact, we end up showing that, under suitable conditions, all the five problems form an equivalence class with respect to the above mentioned relation. 
In this section, we assume $\rho > 0$. First, we make the following observations: 

\begin{itemize}
    \item Among the five problems discussed in the previous section, only Ar{\i}kan's guessing and Huleihel et al.'s memoryless guessing have a unique optimal solution; others only have asymptotically optimal solutions. 
    
    \item Optimal solution of Huleihel et al.'s memoryless guessing problem is the $\alpha$-scaled measure of the underlying probability distribution $P$. Hence, knowledge about the optimal solution of this problem implies knowledge about optimal (or asymptotically optimal) solution of all other problems. 
    
    \item Among the Bunte and Lapidoth's and ordered tasks problems, asymptotically optimal solution of one yields that of the other. The partitioning lemma (Prop. III-2 of \cite{Bunte}) is the key result in these two problems as it guarantees the existence of the asymptotically optimal partitions in both these problems.
\end{itemize}

\subsection{Campbell's Coding and Ar{\i}kan's Guessing}
\label{subsec:codeguess}
 An attempt to find a close relationship between these two problems was made by Hanawal and Sundaresan  \cite[Sec.~II]{HanSunTIT2011}. Here, we show the equivalence between asymptotically optimal solutions of these two problems.

\begin{proposition}
An asymptotically optimal solution exists for Campbell's source coding problem if and only if an asymptotically optimal solution exists for Ar{\i}kan's guessing problem.
\end{proposition}

\begin{IEEEproof}
Let $\{G_n^*\}$ be an asymptotically optimal sequence of guessing functions, that is, $$\lim_{n\to\infty} \frac{1}{n\rho}\log \mathbb{E}_{P_n}[G_n^*(X^n)^\rho] = H_{\alpha}(P).$$ Define
\begin{equation}
\label{eqn:guess_induced_measure}
 Q_{G_n^*}(x^n) := c_n^{-1}\cdot {G_n^*(x^n)}^{-1},
\end{equation}
where $c_n$ is the normalization constant. Notice that 
\begin{eqnarray}
\label{eqn:guessing_cn}
 c_n = \sum_{x^n}{G_n^*(x^n)}^{-1}\le 1 + n\ln |\mathcal{X}|.
\end{eqnarray}
Let us now define
\[
L_{G_n^*}(x^n) := \lceil -\log  Q_{G_n^*}(x^n)\rceil.
\]
Then by \cite[Prop.~1]{HanSunTIT2011},
\begin{equation*}
    L_{G_n^*}(x^n)\le \log G_n^*(x^n) + 1 + \log c_n.
\end{equation*}
Hence
\begin{eqnarray*}
2^{\rho L_{G_n^*}(x^n)} &\le& 2^\rho\cdot c_n^\rho\cdot G_n^*(x^n)^\rho\\
&\le& 2^\rho\cdot (1 + n\ln |\mathcal{X}|)^\rho\cdot G_n^*(x^n)^\rho.
\end{eqnarray*}
Thus, we have
\begin{equation*}
\limsup_{n\to\infty}\frac{1}{n\rho}\log \mathbb{E}_{P_n}[2^{\rho L_{G_n^*}(X^n)}] \le \limsup_{n\to\infty} \frac{1}{n\rho}\log \mathbb{E}_{P_n}[G_n^*(X^n)^\rho] = H_{\alpha}(P).
\end{equation*}
We observe that
\begin{align*}
 \sum_{x^n \in \mathcal{X}^n} 2^{-L_{G_n^*}(x^n)} = \sum_{x^n \in \mathcal{X}^n} 2^{-\lceil -\log Q_{G_n^*}(x^n) \rceil} \leq \sum_{x^n \in \mathcal{X}^n} 2^{\log Q_{G_n^*}(x^n)} = 1.   
\end{align*}
Consequently, from \eqref{eq:Campbell_lower_bound}, we have
\begin{equation*}
\liminf_{n\to\infty}\frac{1}{n\rho}\log \mathbb{E}_{P_n}[2^{\rho L_{G_n^*}(X^n)}] \ge H_{\alpha}(P).
\end{equation*}

Thus, $\{L_{G_n^*}\}$ is an asymptotically optimal sequence of length functions for Campbell's coding problem.

\noindent
Conversely, given an asymptotically optimal sequence of length functions $\{L_n^* \}$ for Campbell's coding problem, define
\[
Q_{L_n^*}(x^n) := \frac{2^{-L_n^*(x^n)}}{\sum_{y^n}2^{-L_n^*(x^n)}}.
\]
Let $G_{L_n^*}$ be the guessing function on $\mathcal{X}^n$ that guesses according to the decreasing order of $Q_{L_n^*}$-probabilities. Then by \cite[Prop.~2]{HanSunTIT2011},
\[
\log G_{L_n^*}(x^n)\le L_n^*(x^n).
\]
Thus
\[
\limsup_{n\to\infty}\frac{1}{n\rho}\log \mathbb{E}_{P_n}[G_{L_n^*}(X^n)^\rho]\le \lim_{n\to\infty}\frac{1}{n\rho}\log \mathbb{E}_{P_n}[2^{\rho L_n^*(X^n)}] = H_\alpha(P).
\]
Further, from Theorem 1 of \cite{Arikan}, we have
\[
\liminf_{n\to\infty}\frac{1}{n\rho}\log \mathbb{E}_{P_n}[G_{L_n^*}(X^n)^\rho]\ge H_\alpha(P) - \limsup_{n\to\infty}\frac{1}{n}\log(1+n\ln |\mathcal{X}|) = H_{\alpha}(P).
\]
This completes the proof.
\end{IEEEproof}

\subsection{Ar{\i}kan's Guessing and Bunte and Lapidoth's Tasks Partitioning Problem}
\label{subsec:guesstasks}

Bracher et al. found a close connection between Ar{\i}kan's guessing problem and Bunte and Lapidoth's tasks partitioning problem in the context of distributed storage \cite{201911TIT_BraHofLap}. In this section, we establish a different relation between these problems.
\begin{proposition}
\label{prop:guess_implies_tasks}
An asymptotically optimal solution of Ar{\i}kan's guessing problem gives rise to an asymptotically optimal solution of tasks partitioning problem.
\end{proposition}
\begin{IEEEproof}
Let $\{G_n^*\}$ be an asymptotically optimal sequence of guessing functions. Define $$Q_{G_n^*}(x^n)  = d_n^{-1} G_n^*(x^n)^{-1/\alpha}, $$
where $d_n$ is the normalization constant. 
Let $A_{G_n^*}$ be the partition function satisfying $A_{G_n^*}(x^n)\le \lceil \beta_n\cdot Z_{Q_{G_n^*}, \alpha}/ Q_{G_n^*}(x^n)^{\alpha}\rceil$ guaranteed by \cite[Prop.~III-2]{Bunte}, where
\[
\beta_n = \frac{2}{M_n-n\log |\mathcal{X}|-2}\cdot
\]
Thus, we have
\begin{eqnarray*}
A_{G_n^*}(x^n)^{\rho} &\le& \lceil \beta_n\cdot Z_{Q_{G_n^*}, \alpha} / Q_{G_n^*}(x^n)^{\alpha}\rceil^{\rho} \\
&\overset{(a)}{\le}& \lceil\beta_n\cdot (1 + n\ln |\mathcal{X}|)\cdot {G_n^*(x^n)}\rceil^{\rho} \\
&\overset{(b)}{\le}& 1 + 2^\rho\beta_n^\rho\cdot (1 + n\ln |\mathcal{X}|)^\rho\cdot {G_n^*(x^n)}^\rho,
\end{eqnarray*}
where (a) holds because $Z_{Q_{G_n^*}, \alpha} = d^{-\alpha}_n \sum_{x^n \in \mathcal{X}^n} {G_n^*}(x^n)^{-1} \leq d^{-\alpha}_n (1 + n \ln |\mathcal{X}|)$; and (b) hold because $\lceil x\rceil^\rho\le 1 + 2^\rho x^\rho$ for $x>0$. Hence
\begin{eqnarray*}
\mathbb{E}_{P_n}[A_{G_n^*}(X^n)^\rho] & \le & 1 + 2^\rho\beta_n^\rho\cdot (1 + n\ln |\mathcal{X}|)^\rho\cdot \mathbb{E}_{P_n}[G_n^*(X^n)^\rho]\\
& \overset{(c)}{\le} & 1 + 2^{2\rho} \left(\frac{1 + n\ln |\mathcal{X}|}{M_n-n\log |\mathcal{X}|-2}\right)^\rho\cdot 2^{n\rho H_\alpha(P)} \\
&{=} & 1 +  2^{n\rho \left(H_\alpha(P) + \frac{\log(1+n \ln |\mathcal{X}|)}{n} - \frac{\log \tilde{M}_n}{n} \right)},
\end{eqnarray*}
where $\tilde{M}_n$ is as in \eqref{eq:mn_tilde}; and inequality (c) follows from Proposition 4 of \cite{Arikan} proved in \cref{sec:problem_statements}. Thus, if $M_n$ is such that $M_n \geq n \log |\mathcal{X}| + 3$ and if $\gamma:= \lim_{n\to\infty} {(\log M_n)}/{n}$ exists and $\gamma > H_{\alpha}(P)$, then we have
\[
\limsup_{n\to\infty}\mathbb{E}_{P_n}[A_{G_n^*}(X^n)^\rho] \le 1.
\]
Since $\mathbb{E}_{P_n}[A_{G_n^*}(X^n)^\rho] \ge 1$, we have $\liminf_{n \to \infty} \mathbb{E}_{P_n}[A_{G_n^*}(X^n)^\rho] \geq 1$. When $\gamma < H_{\alpha}(P)$, arguing along the lines of proof of \cref{thm:tasks_partition_refined}(b), it can be shown than 
$$\lim_{n\to\infty} \frac{1}{n \rho} \log \mathbb{E}_{P_n}[A_{G_n^*}(X^n)^\rho]  = H_{\alpha}(P) - \gamma.$$
\end{IEEEproof}

Reverse implication of the above result does not hold always due to the additional parameter $M_n$ in the tasks partitioning problem. For example, if $M_n = |\mathcal{X}|^n$ and $A_n(x^n) = 1$ for every $x^n$, the partition does not provide any information about the underlying distribution. As a consequence, we will not be able to conclude anything about the optimal (or asymptotically optimal) solutions of other problems. However, if $M_n$ is such that $\log M_n$ increases  sub-linearly, then it does help us find asymptotically optimal solutions of other problems.

\begin{proposition}
\label{thm:connection_partition_guessing}
An asymptotically optimal sequence of partition functions $\{A_n\}$ with partition sizes $\{ M_n \}$ for the tasks partitioning problem gives rise to an asymptotically optimal solution for the guessing problem if $M_n \geq n \log |\mathcal{X}| + 3$ and 
$\lim_{n\to\infty} {(\log M_n)}/{n} = 0.$
\end{proposition}
\begin{IEEEproof}
By hypothesis, 
\[
\lim_{n\to\infty}\frac{1}{n\rho}\log \mathbb{E}_{P_n}[A_n(X^n)^\rho] = H_{\alpha}(P).
\]
For every $A_n$, define the probability distribution
\begin{equation*}
    Q_{A_n}(x^n) := c_n^{-1} A_n(x^n)^{-1},
\end{equation*}
where $c_n := \sum_{x^n}A_n(x^n)^{-1} = M_n$. Let $G_{A_n}^*$ be the guessing function that guesses according to the decreasing order of $Q_{A_n}$-probabilities. Then by \cite[Prop.~2]{HanSunTIT2011}, we have
\begin{eqnarray*}
G_{A_n}^*(x^n)  \le & Q_{A_n}(x^n)^{-1} = c_n A_n(x^n) = M_n A_n(x^n).
\end{eqnarray*}
Hence
$$
\limsup_{n \to \infty} \frac{1}{n\rho}\log \mathbb{E}_{P_n}[G_{A_n}^*(X^n)^\rho] \le \limsup_{n \to \infty}\frac{1}{n}\log M_n + \limsup_{n \to \infty} \frac{1}{n\rho} \log \mathbb{E}_{P_n}[A_n(X^n)^\rho] = H_{\alpha}(P).
$$
Further, an application of Theorem~1 of \cite{Arikan} gives us
$$\liminf_{n \to \infty} \frac{1}{n\rho} \log \mathbb{E}_{P_n}[G_{A_n}^*(X^n)^\rho] \geq H_{\alpha}(P) - \limsup_{n \to \infty} \frac{1}{n} \log(1+ n \ln |\mathcal{X}|) = H_{\alpha}(P).$$
This completes the proof.
\end{IEEEproof}

\subsection{Huleihel et al.'s Memoryless Guessing and Campbell's Coding}

We already know that if one knows the optimal solution of Huleihel et al.'s memoryless guessing problem, that is, the $\alpha$-scaled measure of the underlying probability distribution $P$, one has knowledge about the optimal (or asymptotically optimal) solution of Campbell's coding problem. In this section, we prove a converse statement. We first prove the following lemma.

\begin{lemma} \label{subsec:lemma_memoryless_coding}
Let $L_n^*$ denote the length function corresponding to an optimal solution for Campbell's coding problem on the alphabet set $\mathcal{X}^n$ endowed with the product distribution $P_n$. Then, 
$\sum \nolimits_ {x^n  \in \mathcal{X}^n} 2^{-L_n^*(x^n)} \geq 1/2$.
\end{lemma}
\begin{IEEEproof}
Suppose $\sum_ {x^n  \in \mathcal{X}^n} 2^{-L_n^*(x^n)} < 1/2$. Then, we must have $L_n^*(x^n) \geq 2$ for every $x^n \in \mathcal{X}^n$. Define $\hat{L}_n(x^n) := L_n^*(x^n) - 1$. We observe that $\sum_ {x^n  \in \mathcal{X}^n} 2^{-\hat{L}_n(x^n)} < 1$, that is, the length function $\hat{L}_n(\cdot)$ satisfies \eqref{eq:KraftMcmilan}. Hence, there exists a code $C_n$ for $\mathcal{X}^n$ such that $L(C_n(x^n)) = \hat{L}_n(x^n)$. Then for $\rho > 0$, we have $\log\mathbb{E}_{P_n}[2^{\rho L^{*}_n(X^n)}] > \log\mathbb{E}_{P_n}[2^{\rho \hat{L}_n(X^n)}]$ --- a contradiction.
\end{IEEEproof}

\begin{proposition} \label{subsec:prop_memoryless_coding}
An asymptotically optimal solution for Huleihel et al.'s memoryless guessing problem exists if an asymptotically optimal solution exists for Campbell's coding problem.
\end{proposition}
\begin{IEEEproof}
Let $\{L_n^* , n \geq 1\}$ denote a sequence of asymptotically optimal length functions of Campbell's coding problem, that is, 
\begin{align}
\lim_{n \to \infty} \frac{1}{n \rho} \log \mathbb{E}_{P_n}[2^{\rho L^*_n(X^n)}] = H_{\alpha}(P). \label{eq:ln_opt_cond}
\end{align}
Let us define
$$
Q_{L^*_n}(x^n) := \frac{2^{-L_n^*(x^n)/\alpha}}{\sum_{{\bar{x}^n} \in \mathcal{X}^n }2^{-L_n^*(\bar{x}^n)/\alpha}}.$$
Then, we have
\begin{align*}
I_\alpha (P_n,Q_{L^*_n}) &= 
\frac{\alpha}{1-\alpha}{\rm log}\Big ( \sum\nolimits_ {x^n \in \mathcal{X}^n} P_n(x^n)
\Big [  \sum\nolimits_ {\hat{x}^n \in \mathcal{X}^n}
\Big ( \frac{Q_{L^*_n}(\hat{x}^n)}{Q_{L^*_n}(x^n)}\Big )^\alpha \Big ]^\frac{1- \alpha}{\alpha}\Big ) - H_\alpha(P_n) \\
&= \frac{\alpha}{1-\alpha}{\rm log}\Big ( \sum\nolimits_ {x^n \in \mathcal{X}^n} P_n(x^n)
\Big [  \sum\nolimits_ {\hat{x}^n \in \mathcal{X}^n}
 \frac{2^{-L_n^*(\hat{x}^n)}}{2^{-L_n^*(x^n)}}  \Big ]^{\rho} \Big ) - n H_\alpha(P) \\
 &= \frac{1}{\rho} \log \mathbb{E}_{P_n}[2^{\rho L^*_n(X^n)}] + \log \zeta_n - n H_\alpha(P),
\end{align*}
where $\zeta_n = \sum \nolimits_ {\hat{x}^n \in \mathcal{X}^n} 2^{-L_n^*(\hat{x}^{n})}$. Hence
\begin{align*}
\lim_{n \to \infty} \frac{1}{n} I_\alpha (P_n, Q_{L^*_n}) =  \lim_{n \to \infty} \frac{1}{n \rho} \log \mathbb{E}_{P_n}[2^{\rho L^*_n(X^n)}] + \lim_{n \to \infty} \frac{1}{n}  \log \zeta_n - H_{\alpha}(P) \overset{(a)}{=} 0,
\end{align*}
where (a) holds because $\zeta_n \in [\nicefrac{1}{2}, 1]$ (refer \cref{subsec:lemma_memoryless_coding}). If we assume the underlying probability distribution to be $Q_{L^*_n}$ instead of $P_n$, and perform memoryless guessing according to the escort distribution of $Q_{L^*_n}$, namely $\widehat{Q}^{*}_n(x^n) = Q_{L^*_n}(x^n)^{\alpha} / Z_{Q_{L^*_n},\alpha}$, due to \cref{prop:memoryless_mismatch}, we have 
$$\lim_{n \to \infty} \frac{1}{n\rho} \log \mathbb{E}_{P_n}\big[{V}_{\widehat{Q}^{*}_n, \, \rho}(X^n) \big] = H_{\alpha}(P) +  \lim_{n \to \infty} \frac{1}{n} I_\alpha (P_n, Q_{L^*_n}) = H_{\alpha}(P).$$
\end{IEEEproof}

\section{Summary and Concluding Remarks} \label{sec:summary}
The main motivation of this paper was the need to unify the source coding, guessing and the tasks partitioning problems by investigating the mathematical and operational commonalities among them. To that end, we formulated a general moment minimization problem and observed that optimal value of its objective function is bounded below by R\'enyi entropy. We then re-established all achievable lower bounds in each of the above-mentioned problems using the generalized framework. It was interesting to note that the optimal solution did not depend on the moment function $\psi$, but only on the underlying probability distribution $P$ and order of the moment $\rho$ (refer \cref{prop:lower_bound_unified}). We also presented a unified framework for the mismatched version of the above mentioned problems. This framework not only led to refinement of the known theorems, but also helped us identify a few new results. We went on to extend the tasks partitioning problem by asking a more practical question and solved it using the proposed unified theory. Finally, we established a close relationship among these problems. 

The established unified framework has the capacity to act as a general toolset and provide insights for a variety of problems in Information Theory. 
For example, we could solve the ordered tasks partitioning problem using this framework. 
The presented unified approach can also be extended and explored further in several ways. This includes, (a) \emph{Extension to general state-space}: It would be interesting to see if the studied problems can be formulated and solved, for example, for countably infinite or continuous support sets. (b) \emph{Applications}: Ar{\i}kan showed an application of the guessing problem in a sequential decoding problem \cite{Arikan}. Humblet showed that
cumulant of code-lengths arises in
minimizing the probability of buffer overflow in source coding problems \cite{Humblet}. Sundaresan \cite{HanS19} and Salamatian et al. \cite{Salman19} show application of memoryless guessing in password attack problems. It would be interesting to see if other potential applications emerge from this unified study.

\bibliographystyle{IEEEtran}
\bibliography{Unification}

\end{document}